\documentclass[aps,prl,reprint,amsmath,amssymb,twocolumn,superscriptaddress]{revtex4}
\usepackage{bbm}
\usepackage{graphicx}
\usepackage{amsmath}
\usepackage{amssymb,amsthm}
\usepackage{color}
\newcommand{\ket}[1]{|#1\rangle}

\newcommand{\bra}[1]{\langle#1|}

\def\eq{\begin{eqnarray}}
\def\en{\end{eqnarray}}
\def\beq{\begin{eqnarray}}
\def\een{\end{eqnarray}}
\def\bfig{\begin{figure}}
\def\efig{\end{figure}}

\begin{document}

\title{Stringent and efficient assessment of Boson-Sampling devices}
\author{Malte C. Tichy}
\affiliation{Department of Physics and Astronomy, Aarhus University, DK-8000 Aarhus, Denmark}
\author{Klaus Mayer}
\affiliation{Physikalisches Institut, Albert-Ludwigs-Universit\"at Freiburg, D-79104 Freiburg, Germany}
\author{Andreas Buchleitner}
\affiliation{Physikalisches Institut, Albert-Ludwigs-Universit\"at Freiburg, D-79104 Freiburg, Germany}
\author{Klaus M{\o}lmer}
\affiliation{Department of Physics and Astronomy, Aarhus University, DK-8000 Aarhus, Denmark}

\begin{abstract} Boson-Sampling  holds the potential to experimentally falsify  the  Extended Church Turing thesis. The computational hardness of Boson-Sampling, however, complicates the   \emph{certification} that an experimental  device yields correct results  in the  regime in which it outmatches classical computers. 
To certify a boson-sampler, one needs to verify quantum predictions and rule out models that yield these predictions without true many-boson interference. We show that a semiclassical model for many-boson propagation reproduces coarse-grained observables that were proposed as witnesses of Boson-Sampling. A   test based on Fourier matrices is demonstrated to  falsify  physically plausible alternatives to coherent many-boson propagation.
\end{abstract}
\date{\today}
\maketitle
\emph{Introduction --} According to the Extended Church Turing thesis (ECT), any efficient computation performed by a physical device can also be performed efficiently (with polynomial overhead) by a classical computer \cite{Aaronson:2013kx}. Since the advent of quantum computation -- especially since the formulation of Shor's factoring algorithm \cite{shor:303,Nielsen:2000fk} -- the ECT has been under attack, since quantum computers are believed to outperform classical devices. Nevertheless, the  available  \emph{empirical} evidence  is  insufficient to dismiss the ECT as a central dogma of computer science, and    a functional universal quantum computer is not likely to be constructed  in the foreseeable future.

A more approachable challenge to the ECT is provided by \emph{Boson-Sampling} \cite{Aaronson:2011kx}, which is  \emph{harder} than factoring, while  it can be solved efficiently by a quantum device of  modest capabilities: Only the coherent propagation of many identical bosons through a multimode setup is required. Experimental boson-samplers with  three photons  match the theoretically expected  particle distribution \cite{Tillmann:2012ys,Crespi:2012vn,Broome15022013,Spring15022013}. Scaling to larger photon numbers is equally challenging \cite{rhode2012,Shchesnovich:2013yq,Leverrier} as conceptually indispensable, and may   be alleviated by alternative 
 formulations of the problem that keep its computational hardness 
  \cite{Lund:2013eb,Rhode:2013vn}. 

The \emph{certification} of an alleged boson-sampler in the regime of many particles is decisive for a serious and well-founded attack on the ECT. Under the assumption that quantum physics correctly describes  the propagation of arbitrarily many bosons, no certification issue arises at all, and no traditionally trained physicist will question the implications of Boson-Sampling.
But in a cross-disciplinary context that encompasses computer science, mathematics and physics, the validity of quantum mechanics for truly many interfering particles must be underpinned by unambiguous empirical evidence.

The very hardness of Boson-Sampling makes such desirable certification a dilemma: On the one hand, it quickly becomes unfeasible to compute  the full Boson-Sampling distribution classically, because the computational expenses for a single transition probability as well as the total number of events diverge exponentially in the number of bosons $n$.
On the other hand, one may measure efficiently predictable observables such as statistical bosonic signatures, but such  strategy  leaves room for alternative models that explain the observed behavior \emph{without} the   complex interference of many bosons. 

The	persuasiveness of any certification protocol  therefore hinges on how convincingly it establishes the quantum prediction for many bosons while ruling out alternative models.  Several efficient certification protocols have been devised \cite{Aaronson:2013ls,Spagnolo:2013eu,Carolan:2013mj,Carolan:2014}; in particular those recently put forward in Ref.~\cite{Carolan:2013mj,Carolan:2014}  discriminate the bosonic output against the classical behavior of distinguishable particles. Here, we show
that certificates based on bosonic bunching are nevertheless
 not \emph{stringent}, because they (erroneously) qualify the output of the efficient 
  and physically plausible semi-classical mean-field sampler 
   (described  below) as a functional boson-sampler. 
We devise an efficient and more stringent test based on highly symmetric sampling matrices, which can conclusively rule out the mean-field sampler and leaves  no room for physically plausible models that pass the test without invoking the granular quantum 
 interference of $n$ bosons. By assessing the  gradual failure of the test due to  inaccuracies, we establish the experimental requirements for a certifiable device.

\begin{figure}[th] \center
\includegraphics[width=\linewidth,angle=0]{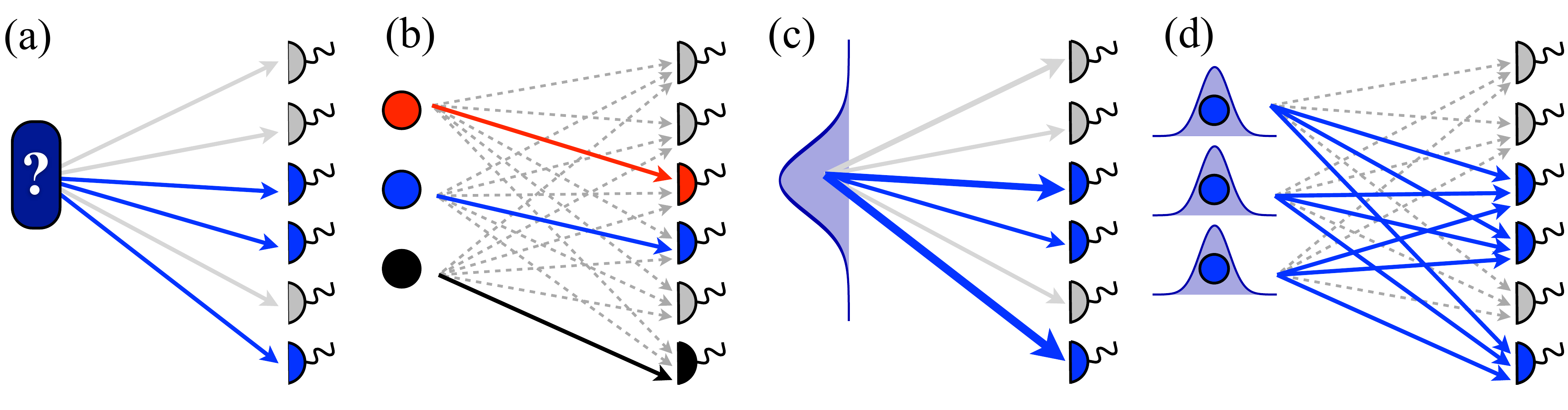}
\caption{(color online) Sampling models. (a) Uniform sampler: The scattering matrix and the initial state are ignored.
(b) Classical sampler: Distinguishable particles propagate independently without interference. (c) Mean-field  sampler: Macroscopic interference and bosonic effects are incorporated. (d)  Boson-Sampling requires  the interference of all $n!$ paths of the $n$-boson wavefunction.
 }  \label{Fig1.pdf}
 \end{figure}

\emph{Sampling and certification -- } Boson-Sampling consists in simulating output events of $n$ bosons  prepared in the $n$ different input ports $\vec j=(j_1, \dots, j_n)$ out of the ${m\gg n}$ modes  of a scattering setup 
 chosen randomly according to the Haar measure on $m\times m$ unitary matrices $U$. That is, one draws output events $\vec k=(k_1, \dots , k_n)$ with probability $P_{\text{B}}( \vec j, \vec k ; U)$, which corresponds to the \emph{permanent} of the sub-matrix of $U$ that contains the rows and columns matching the occupied input and output modes,
\eq
 P_{\text{B}}( \vec j, \vec k ; U)  =  | \text{permanent}(M) |^2 ,  \label{permanenteq} ~~~M_{l,q} &= & U_{j_l, k_q} ,
\en
where  additional combinatorial factors arise for multiply occupied output modes.
 The permanent  eludes  polynomial algorithms, which is inherited by the above sampling problem: Physically speaking, the interference of $n!$  many-particle paths \cite{Tichy:2012NJP} governs each event [see Fig.~\ref{Fig1.pdf}(d)].
 An efficient classical  algorithm for Boson-Sampling would imply extremely surprising consequences in computational complexity theory \cite{Aaronson:2011kx}.

To certify a boson-sampler, one needs to verify predictions following from Eq.~(\ref{permanenteq}) and rule out models that yield these predictions without true many-boson interference \cite{gogolin2013}. 
A certification protocol needs to be discarded if it accepts a series of events produced by a fraudulent device. Relevant fraudulent devices are those based on a \emph{plausible physical} mechanism that can be simulated \emph{efficiently} on a classical \emph{computer}.

 The simplest way to efficiently sample from the space of events $\vec k$ is \emph{uniform sampling} [Fig.~\ref{Fig1.pdf}(a)], for which each
 event is assigned the same probability \cite{gogolin2013}, and no
 information on the matrix $U$ or the initial state $\vec j$ is exploited. When $U$ and $\vec j$ 
  are provided, one can distinguish Boson-Sampling from
a uniform sampler via, e.g., the average number of particles in each output mode \cite{PhysRevA.83.062307,Aaronson:2013ls,Spagnolo:2013eu}, which can be computed for the boson-sampler without evaluating any permanent,
 \eq
  \langle \hat n_k \rangle = \sum_{l=1}^n |U_{j_l, k}|^2  . \label{meanpnum}
 \en

Such \emph{single-particle} observables reflect certain properties of the matrix $U$, but they are insensitive to many-particle interference \cite{PhysRevA.83.062307,Tichy:2012NJP}. 
Therefore, single-particle observables are also replicated by efficient \emph{classical sampling}, which  can be implemented physically with distinguishable particles:
Events are constructed by choosing the output mode $k$ for each particle prepared in $j_q$ with probability $p_{j_q,k}=|U_{j_q,k}|^2$ [see Fig.~\ref{Fig1.pdf}(b)].
Single-particle observables are therefore not sufficient to discriminate the output of a boson-sampler from a classical or a fermion-sampler 
 \cite{PhysRevA.83.062307,Aaronson:2013ls,Spagnolo:2013eu}.

To rule out the classical sampler, 
 appropriate coincidence and correlation observables were proposed in  \cite{Carolan:2013mj,Carolan:2014}: The probability $P_1$ to find an $n$-fold coincidence outcome (without any multiply populated mode) is significantly higher for distinguishable particles than for bosons due to the bunching tendency of the latter. Similarly, one can leave the space of random  matrices and focus on structured multimode setups with certain features: \emph{Bosonic clouding} \cite{Carolan:2013mj} is the tendency for bosons to favor events with  all particles in the same half of the output array of a continuous-time many-particle quantum walk.         
 The two proposed observables, however,  do \emph{not} unambiguously verify the \emph{many-body coherence} of the wavefunction, since the statistical tendency to multiply populate output states survives the deterioration of 
 granular many-body interference: In the \emph{mean-field} sampler, the Wigner function \cite{Dowling1994}  of the multi-mode  Fock-state is semi-classically approximated by the macroscopically populated single-particle states [see Fig.~\ref{Fig1.pdf}(c)] 
\eq
 \ket{\psi} = \frac1  {\sqrt{n}} \sum_{r=1}^n e^{i \theta_r} \ket{\phi^{(\text{in})}_{j_r} } , \label{iniMF}
\en
where the phases $\theta_r$ are undefined \cite{Chuchem2010,Mullin:2010vn,TichyDiss}. That is to say, the mean field  forms a thin belt on the high-dimensional Bloch-sphere \cite{Dowling1994}, which 
 evolves into 
\eq
 \hat U \ket{\psi} & =&  \frac 1 {\sqrt{n}}\sum_{q=1}^m  \left[ \ket{\phi^{(\text{out})}_q} \left( \sum_{r=1}^n e^{i \theta_r}  U_{j_r, q} \right) \right], \label{finalmf}
\en
i.e.~each particle occupies the output mode $1\le q \le m$ with probability $p^{\text{mf}}_q=  |\bra{\phi_q^{(\text{out})} } \hat U \ket{\psi} |^2 = \frac 1 n \left| \sum_{r=1}^n e^{i \theta_r} U_{j_r,q } \right|^2$ \cite{Mullin:2010vn}. The ensemble average consists in sampling over random phases $\{ \theta_1 \dots \theta_n \}$, each setting  then leaves the particles (classically) correlated, since particles gather where  $p_q^{\text{mf}}$  is high.

The mean-field sampler is an efficiently evaluable and physically plausible model: It contains those aspects of many-boson dynamics that survive in the semi-classical limit, in which fragile many-boson quantum interference is lost, and it describes experiments with interfering Bose-Einstein condensates \cite{PhysRevLett.93.180403,Cennini:2005th}. It can be implemented alternatively by sequentially preparing $n$ particles in the same initial state (\ref{iniMF}).

The mean-field sampler yields the expected mean occupation (\ref{meanpnum}) and, as shown in Fig.~2, it reproduces the coincidences $P_1$ and the clouding $C$ predicted for the boson-sampler, Eq.~(\ref{permanenteq}). This clearly dismisses these observables as witnesses of Boson-Sampling. All coarse-grained signatures that can be ascribed to \emph{bosonic statistics} are reproduced by mean-field sampling and cannot validate the potential of a physical device to disprove the ECT. 
Similar to the request that stringent tests of entanglement rule out behavior borne by classical fields with random correlated phases \cite{PhysRevLett.88.097902}, a certification protocol for boson-sampling must include tests which address properties that are not reproduced by the mean-field  sampler.

\begin{figure}[ht] \center
\includegraphics[width=.8\linewidth,angle=0]{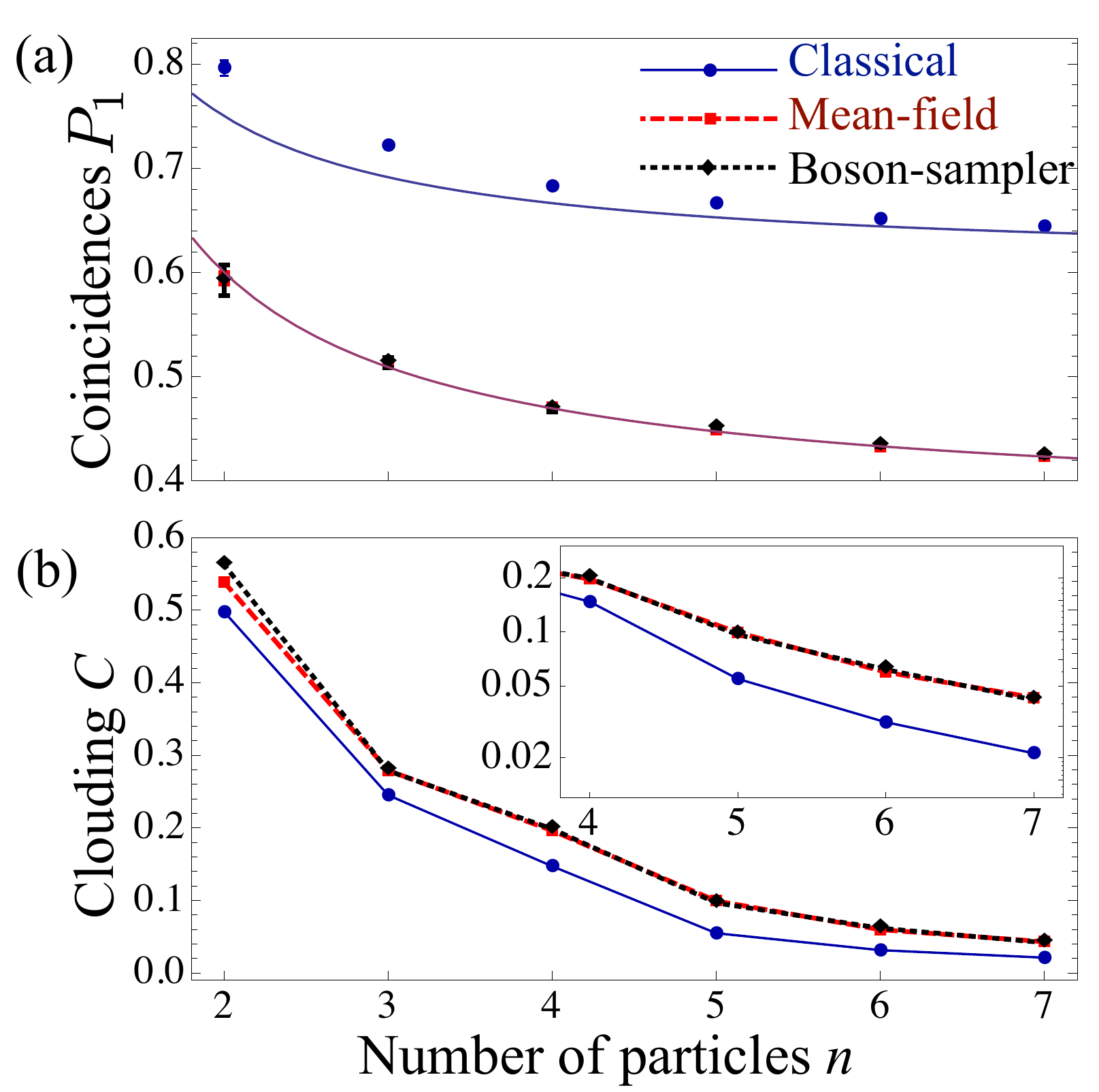}
\caption{(color online) Bunching and clouding in different sampling models. Classical distinguishable particles (blue circles) are compared to the mean-field sampler (red squares) and the boson-sampler (black diamonds).  (a) Coincidence probability $P_1$ for 100 Haar-random unitary matrices of dimension $m=n^2$ with error-bars that represent one standard deviation (the mean-field sampler is hardly discernible from the boson-sampler). The lines are the combinatorially expected probabilities \cite{Carolan:2013mj}.
(b) Clouding for a discrete-time quantum walk of 8 steps with $n$ particles starting in adjacent modes \cite{PhysRevA.83.062307,mayerda}. The probability $C$ that all particles be in the same half of the output array coincides for the mean-field  and the boson-sampler. 
}
\label{Fig2CloudingAndBunching.pdf}
 \end{figure}

\emph{Certification via Fourier matrices --} A certification scheme which rules out plausible physical models that circumvent the evaluation of the permanent (\ref{permanenteq}) needs to be based on efficiently predictable, fine-grained observables that are sensitive to granular $n$-body interference. Since event probabilities subjugated by $n$-body interference are hard to predict for unstructured random matrices, we leave the space of computationally hard sampling problems and choose a physically non-trivial, albeit efficiently predictable artificial \emph{instance} of Boson-Sampling: The difficulty in the evaluation of the permanent in Eq.~(\ref{permanenteq}) in comparison to the benevolent determinant is due to the lack of symmetries. In order to alleviate the complexity, we  choose a  symmetric sampling matrix,  the
 Fourier matrix of dimension $m=n^p$,
\eq
 U^{\text{Fou}}_{l,q}=\frac{1}{\sqrt{m}} \text{exp}\left(i \frac{2 \pi l q }{ m} \right) . \label{fouriermatrix}
\en
Cyclic symmetry is imposed on the initial state,
\eq
 \vec j_{\text{cyc}} = \left(1, n^{p-1}+1, 2 n^{p-1} +1, \dots , (n-1) n^{p-1} +1 \right) . \label{inistatecyclic}
\en
The cyclic symmetry remains intact during the scattering process, which is reflected by the occurring output events: Many  events $\vec k$ are \emph{forbidden} due to the
 suppression law for Fourier matrices \cite{Tichy:2010ZT,Tichy:2012NJP,TichyDiss}:
\eq
 \text{mod}\left( \sum_{l=1}^{n} k_l  , n  \right) \neq 0  \Rightarrow  P_{\text{B}}(\vec j_{\text{cyc}}, \vec k; U^{\text{Fou}} )=0 , \label{conditioin}
\en
which generalizes the Hong-Ou-Mandel effect: 
For $\vec j=(1,2)$, coincident output events with  $\vec k=(1,2)$ lead to an odd sum in (\ref{conditioin}), two photons then  never leave the beam-splitter in different modes \cite{Hong:1987mz}.

\begin{figure}[ht] \center
\includegraphics[width=.7\linewidth,angle=0]{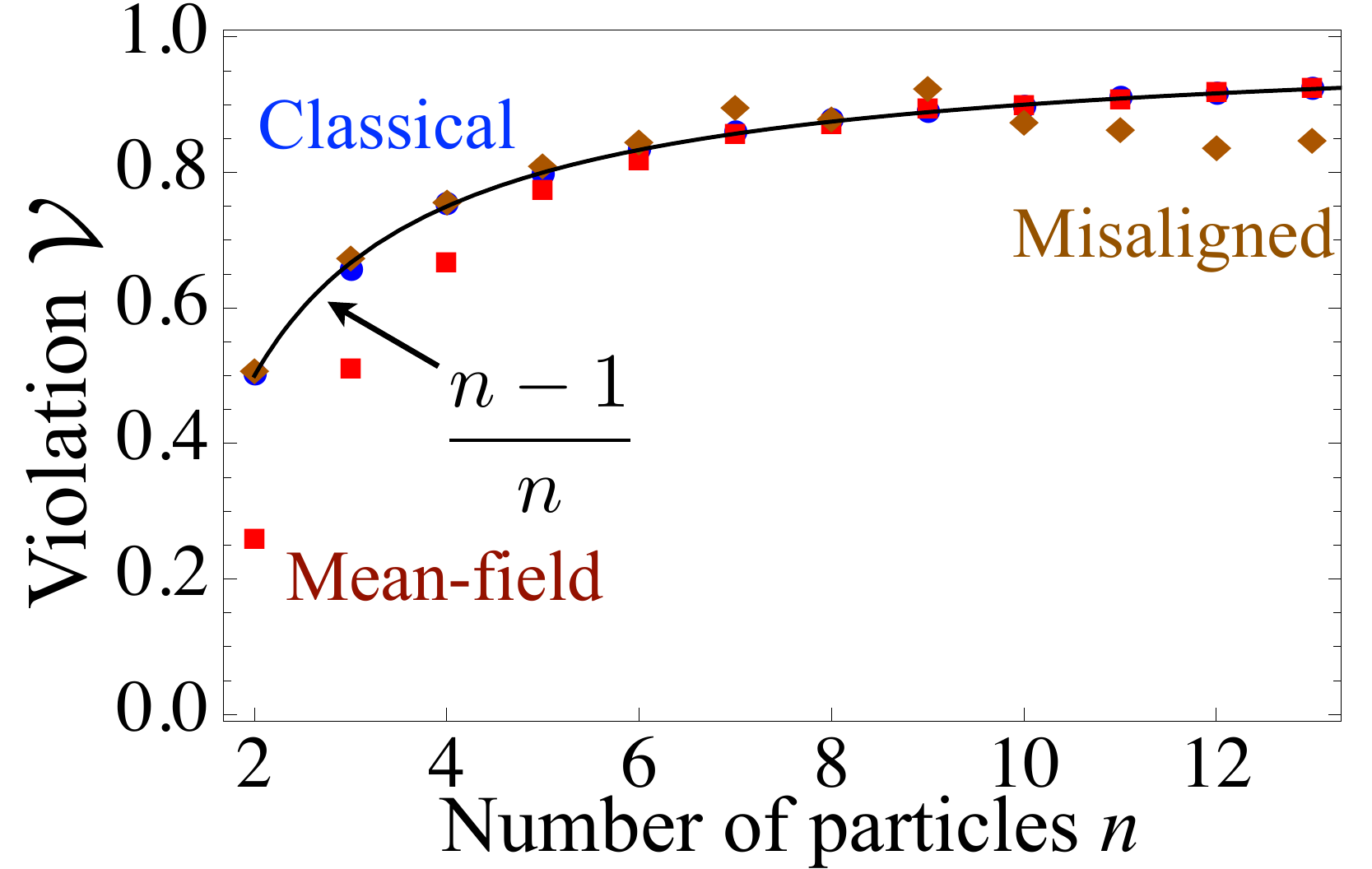}
\caption{(color online) Violation $\mathcal{V}$  of the suppression law (\ref{conditioin}) by the classical (blue circles), mean-field  (red squares) and misaligned (brown diamonds) sampler for $m=n^2$.
The classical violation coincides with the ratio of suppressed events for the boson-sampler, 
 $(n-1)/n$ (black solid line). For small particle numbers $n \lessapprox 4$, the suppression law favors bunched states with many particles in few modes, which alleviates 
 the violation by the mean-field  sampler. 
For the misaligned data, we assume 
 that one particle out of $n$ is distinguishable from the others, the total violation is inferred from computing the probability for up to 10000 distinct forbidden output states.  }  \label{Fig2aviolationSL.pdf} 
 \end{figure}

The degree of violation of the suppression law is quantified by  $\mathcal{V}={ \mathcal{N}_{\text{forbidden}}   }/ {\mathcal{N}_{\text{runs} }} $, the ratio of actually occurring events ${\mathcal{N}_{\text{forbidden}}}$ that violate Eq.~(\ref{conditioin}) to the total number of events $\mathcal{N}_{\text{runs}} $.
An ideal  boson-sampler features $\mathcal{V}=0$. The uniform, classical and mean-field samplers do not contain any mechanism to satisfy  the suppression law, which leads to a considerable violation:
Only a fraction $1/n$ of a priori possible events  can occur in an accurate Boson-Sampling experiment on the Fourier matrix (\ref{fouriermatrix}) and the initial state (\ref{inistatecyclic}), while \emph{most events are forbidden} \cite{Tichy:2010ZT}.
For uniform, classical and  mean-field sampling, one therefore observes $\mathcal{V} \approx (n-1)/n$ for large $n$ [see Fig.~\ref{Fig2aviolationSL.pdf}], and the probability that all events out of a sample of  $R$ events accidentally fulfil the suppression law is $1/n^R$. The Fourier matrix does not constitute a \emph{complex} scenario, since forbidden events can be predicted efficiently: In practice,  
merely fractions of a second are required on a PC for $n \sim 10^6$. Notwithstanding, the observation of the suppression law in an experiment relies on granular $n$-particle interference: All $n!$ many-particle amplitudes need to perfectly cancel, making the method \emph{stringent}.

 Can other models for many-body propagation fulfil the suppression law \emph{without} true many-boson interference inherent to Eq.~(\ref{permanenteq})? From the \emph{computational} point of view, there are efficient fraudulent models: The output of a mean-field sampler can be checked against the suppression law, and forbidden events are blocked (alternatively, for odd $n$, bosonic forbidden events are also forbidden for efficiently simulatable fermions \cite{Tichy:2012NJP}, although the latter do not bunch).  From a \emph{physical} perspective, however, there is no plausible mechanism that reads out the artificial symmetries (\ref{fouriermatrix}), (\ref{inistatecyclic}) of the setup, establishes the suppression law (\ref{conditioin}), and implements an ad-hoc veto on the output states: The suppression of an event is a collective non-local property of the output state $\vec k$, which requires a physical mechanism that reigns over all particles in a concerted way. Hence, fulfilling the suppression law qualifies as the desired  convincing ``circumstantial evidence'' \cite{Aaronson:2013ls} that an alleged boson-sampler is operational.

\emph{Deterioration due to inaccuracies --} Our criterion based on the suppression law might  appear too stringent in practice: Deviations from the ideal can be expected due to experimental inaccuracies, such as partial distinguishability of the bosons \cite{TichyFourPhotons,Ra:2013kx,younsikraNatComm,Tillmann2014} and  deviations of the scattering matrix $U$ from the Fourier matrix (\ref{fouriermatrix}).
The prediction of individual event probabilities in such scenario  is unfeasible for many particles  \cite{TichyFourPhotons}, but the large fraction $(n-1)/n$ of forbidden events allows us to efficiently estimate the violation  $\mathcal{V}$, as shown in the following.

A state of partially distinguishable bosons reads \cite{TichyFourPhotons}
\eq
 \ket{\Psi^{\text{ini}}} =\prod_{r=1}^n \hat a^\dagger_{j_r, t_r} \ket{\text{vac}} ,
\en
where the distinguishing degree of freedom $t_r$ accounts for, e.g., the mutual delay  of injected photons. The states $\ket{t_1}, \dots \ket{t_n}$ are Gram-Schmidt-orthogonalized to give the orthonormal basis $\{ \ket{t_1}, \ket{\tilde t_2} \dots \ket{\tilde t_n} \}$, which permits to expand $\ket{\Psi^{\text{ini}}}$ into 
 $n!$ orthogonal terms   \cite{TichyFourPhotons,Ra:2013kx,younsikraNatComm},
\eq
 \ket{\Psi^{\text{ini}}} = \hat a^\dagger_{j_1, t_1 }  \sum_{d_2=1}^2 \sum_{d_3=1}^3 \dots \sum_{d_n=1}^n   \prod_{r=2}^n c_{r,d_r} \hat a^\dagger_{j_r, \tilde t_r} \ket{\text{vac}}  . \label{orthonormalized}
\en
Each summand describes a different degree of interference capability:
The term weighted by $c_{2,1}c_{3,1}\dots c_{n,1}$ with $d_{2\dots n}=1$ describes  indistinguishable bosons that interfere perfectly and only give rise to non-forbidden events ($\mathcal{V}=0$). The term with $d_q=q~(2 \le q \le n)$ describes  distinguishable particles, which induces $\mathcal{V} \approx (n-1)/n$. 
 Intermediate terms that describe neither fully distinguishable nor fully indistinguishable particles give rise to bosonic signatures such as bunching, but, in most cases, they do not fulfil the suppression law, and induce a violation of the order $(n-1)/n$. Even when merely one out of $n$ bosons is distinguishable, the suppression law is strongly violated [see brown diamonds in Fig.~\ref{Fig2aviolationSL.pdf}]. The total  distinguishability-induced violation $\mathcal{V}_{\text{partial}}$ is therefore bounded by the weight of the perfectly indistinguishable term,
\eq
 \mathcal{V}_{\text{partial}} \lessapprox  \frac{n-1}{n} \left(1 - \prod_{q=2}^n |c_{q,1}|^2 \right) .
\en

\begin{figure}[ht] \center
\includegraphics[width=.8\linewidth,angle=0]{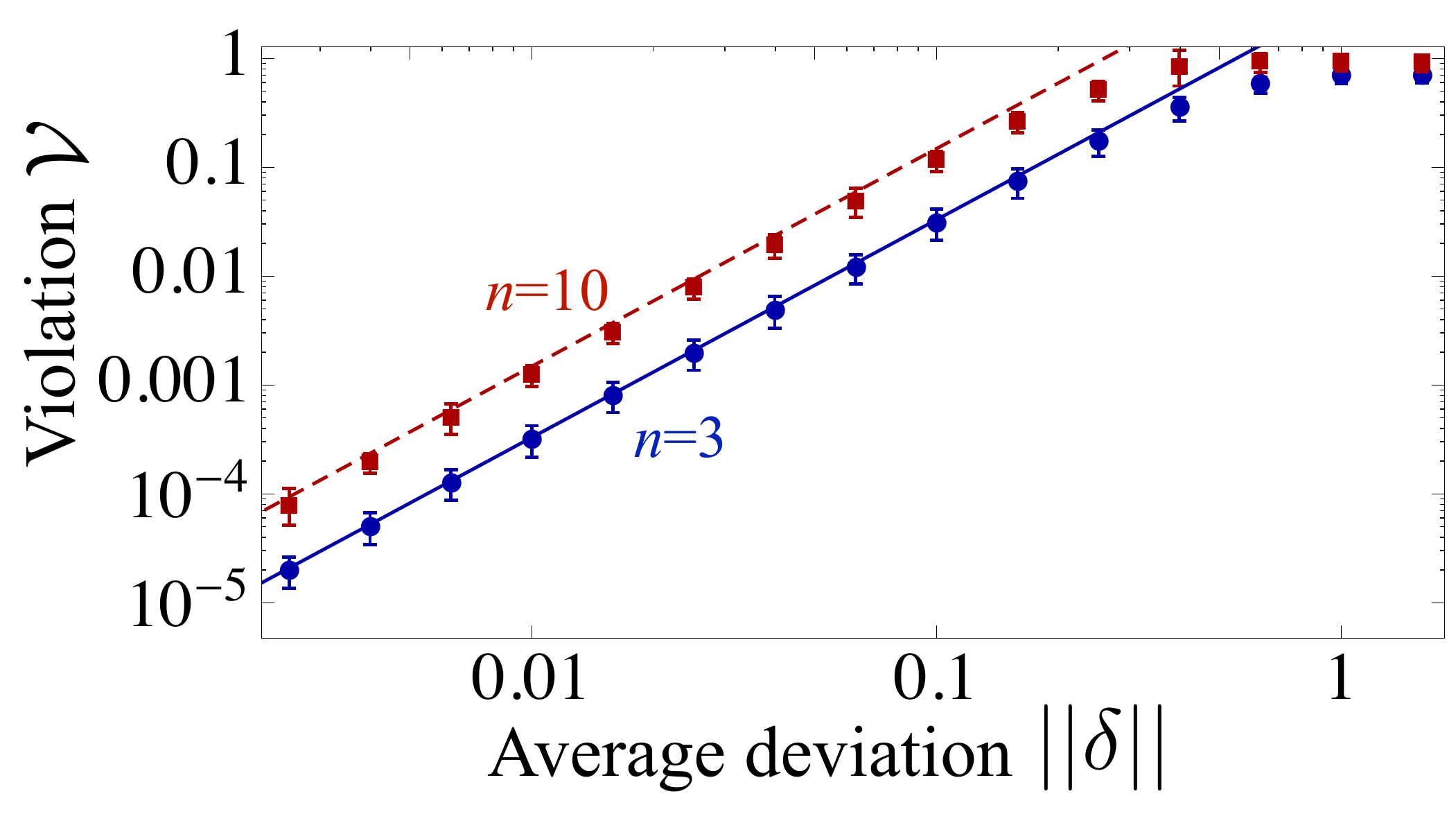}
\caption{(color online) Violation $\mathcal{V}$ for matrices described by Eq.~(\ref{Wmatrix}), for $n=3,10$ and $m=n^2$. We compute the total probability of 200 randomly chosen forbidden events, for 400 different matrices $\delta$ for each value of the average deviation
 $|| \delta ||$.   Error bars represent one standard deviation, the dashed red and solid blue lines show  the estimate  (\ref{esimt2}) for $n=10$ and $n=3$, respectively, which breaks down when $||\delta||$ is not  smaller than $1/n$. In order to observe the suppression law in the experiment, the violation needs to be significantly smaller than $(n-1)/n$ (compare to Fig.~\ref{Fig2aviolationSL.pdf}). }  \label{Fig3est.pdf}
 \end{figure}

Another experimental limitation is that the desired unitary transformation can  be implemented only with limited accuracy. The probability for a forbidden transition $\vec j \rightarrow \vec k$ is not described by a submatrix $M$ of the Fourier matrix $U^{\text{Fou}}$ [see Eq.~(\ref{fouriermatrix})], but by a matrix $W$ with
\eq
 W_{l,q} = M_{l,q}\left( 1 + \delta_{l,q} \right)  . \label{Wmatrix}
\en
By expanding the permanent of $W$ in powers of matrix elements of $\delta$ to the first order, we can estimate
 \eq
 |\text{permanent}( W ) |^2 &\approx & P_{\text{approx}}(\delta) :=  \frac{ n \cdot n!  }{m^n}  ||  \delta  ||^2   \label{perest} ,
 \en
  where $|| \delta || := \langle |  \delta_{l,q} | \rangle_{l,q}$ is the average absolute value of matrix elements of $\delta$, which we assume to be much smaller than unity.
For small deviations,  the probability for non-forbidden events remains widely unaffected by $\delta$, and -- circumventing the permanent -- the violation  can be estimated as
\eq
 \mathcal{V}_{\text{dev}} &\approx&  \frac{n-1}{n} {N}_{\text{events}}  P_{\text{approx}}( \delta ) , \label{esimt}  \\ &\stackrel{\text{for }m=n^2}{\approx } & \sqrt{e}~ (n-1)  || \delta ||^2 ,\label{esimt2} 
\en
where  $ {N}_{\text{events}}= {m+n-1 \choose n}$ is the total number of events.
The estimate  is confirmed numerically in Fig.~\ref{Fig3est.pdf} for ${n=3,10}$, $m=n^2$.
Eqs.~(\ref{esimt}), (\ref{esimt2}) also formalize the mild requirement on the  accuracy of multimode scattering matrices that feature the suppression law. Since the two sources of deterioration are independent, the total expected violation 
 can be estimated as $ \mathcal{V}_{\text{total}}  \approx    \mathcal{V}_{\text{partial}}  +  \mathcal{V}_{\text{dev}}  $.

\emph{Outlook --}
The potential influence of a boson-sampler on the foundations of theoretical computer science is rather formidable, but so will be the requirements on convincing
 evidence for its proper functionality. Therefore, high exigency needs to be imposed on the  falsification of alternative models for many-particle behavior. We showed that coarse-grained criteria based on bosonic bunching or clouding \cite{Carolan:2013mj,Carolan:2014} are insufficient, since they are reproduced by the semi-classical mean-field sampler 
  \cite{PhysRevLett.93.180403,Cennini:2005th}.

A functional boson-sampler will necessarily implement any unitary sampling matrix that the user wishes for to switch between different instances of the problem \cite{Reck}, and we can focus on the special instance described by (\ref{fouriermatrix}),  (\ref{inistatecyclic}) to assess many-particle interference. 
The verification of (\ref{conditioin}) with three photons \cite{Spagnolo:2013fk} and  progress in integrated waveguide techniques \cite{Spagnolo:2012kn,Meany:12,J2012ml} feed the hope that the suppression law (\ref{conditioin}) will be observed in more complex  setups in the near future. Within quantum mechanics, there are two sources of deviation from the ideal: Bosons can carry distinguishing degrees of freedom, and the setup might not precisely match the Fourier matrix. The  deterioration induced by both effects can be estimated efficiently.

 Following the spirit of the falsification of local realism 
  \cite{Bell:1964pt}, 
  one may envisage a 
   matrix similar to Eq.~(\ref{fouriermatrix}), but with \emph{hidden} symmetries, such that 
    events are forbidden according to an intricate
   rule that encodes the solution to a computationally hard problem.   While sampling 
    should therefore remain hard to \emph{perform}, the output should be nevertheless easy to \emph{verify} (in the language of computational complexity, the problem encoded by the matrix is in the complexity class NP). Such -- admittedly speculative \cite{Aaronson:2011kx,Aaronson:2013ls} --  instance of asymmetric complexity may offer an unquestionable \emph{computational} criterion for the certification of boson-samplers, and promote such devices into powerful tools for algorithmic applications.

{\emph{Acknowledgements.} M.C.T. would like to thank Scott Aaronson, Christian Gogolin,  Robert Keil and Karol \.Zyczkowski for very helpful and clarifying  discussions. M.C.T and K. M{\o}lmer would like to thank the Villum foundation and the Danish Council for Independent Research.


\end{document}